\documentstyle[12pt,epsf]{article}
\newcommand{\be}{\begin{equation}}
\newcommand{\ee}{\end{equation}}
\newcommand{\ba}{\begin{eqnarray}}
\newcommand{\ea}{\end{eqnarray}}
\newcommand{\bb}{\begin{thebibliography}}
\newcommand{\eb}{\end{thebibliography}}
\newcommand{\ci}[1]{\cite{#1}}
\newcommand{\bi}[1]{\bibitem{#1}}
\newcommand{\lab}[1]{\label{#1}}

\begin{document}
\phantom{.}
\vspace{2cm}
\begin{flushright}
hep-ph 9604261
\end{flushright}
\begin{center}
{\large{Diffractive Contribution to $g_1$ Structure Function at Small $x$}}
\\
S.V.Goloskokov, \footnote{Email: goloskkv@thsun1.jinr.dubna.su}\\
Bogoliubov Laboratory of Theoretical  Physics,\\
 Joint Institute for Nuclear Research,\\
Dubna 141980, Moscow region, Russia
\end{center}

\vspace{.7cm}
\begin{abstract}
We study the small $x$ diffractive contribution to the spin--dependent
proton structure function $g_1(x)$. We find that the $x\to 0$ behaviour of
$ g_1(x)$ has a singular form like $1/(x^\alpha \ln^2(x))$ with $\alpha
\sim 0.2-0.3$.\\ PACS, 12.38.Bx, 13.60.Hb, 13.88.+e
\end{abstract}
\newpage

Polarized structure functions in deep--inelastic scattering
have   recently been studied in several experiments \ci{1,2,smc}. The
data on the $g_1$ structure function provide important information on
spin--dependent parton distributions. The first moment of the $g_1$
function
\be
\Gamma_1= \int_{0}^{1} dx\; g_1(x)                     \lab{gam}
\ee
is used to test the Ellis--Jaffe \ci{ej} and Bjorken \ci{bj} sum--rules.
The detailed discussion of different aspects of the theory and
phenomenology of polarized deep--inelastic scattering can be found in
\ci{ans}.

One of the milestones in calculation of $\Gamma_1$ (\ref{gam}) is the
extrapolation of the $g_1(x)$ structure function to the $ x\to 0$ region.
The small--$x$ behaviour of $g_1$ is not understood very well now. The
theoretical study of this problem should be important for future spin
experiments at HERA where the test of the  $x \geq 10^{-4}$ region can be
made \ci{blu}.

There are different parametrizations that show either positive
\ci{gp} or negative \ci{gn} behaviour of $g_1$ for $x \le 10^{-3}$.
The Regge theory leads to
\be
g_1(x) \sim x^\alpha                  \lab{alf}
\ee
with $\alpha=\alpha_{A_1}$ ($0 \leq \alpha \leq 0.5$) determined by the
contribution of the $A_1$ trajectory \ci{a1}. This form is usually used by
experimental groups to determine  the  low--$x$ behaviour of $g_1$.
The standard extrapolation for $g_1$  is  $g_1 \sim 0.35$ for $x \le
0.01$.

For the first time the SMC experiment \ci{smc} has indicated
possible growth of $g_1^p$ at small $x$. The singular behaviour of $g_1$
can be found by studying the double--logarithmic contributions of
perturbative QCD \ci{dln,dln1}. Using QCD arguments the form of $g_1$ like
(\ref{alf}) with $\alpha=1-\alpha_{P}(0)$ ($\alpha_{P}(0)$ is the pomeron
intercept) can be obtained \ci{dln}.

The model of the pomeron made up of nonperturbative gluons \ci{lana}
has been used to examine the $g_1$ structure function  at small $x$
\ci{land}.  The following form was obtained:
\be
g_1^L(x)=0.09(2 \ln(\frac{1}{x})-1).                \lab{g1l}
\ee
It was found that the pomeron-pomeron cut contribution to $g_1$ could
behave as \ci{clo}
\be
g_1^C(x)  =0.17 \frac{1}{x \ln^2{x}}.                \lab{g1c}
\ee
The models (\ref{g1l},\ref{g1c}) are  consistent with the SMC data on
$g_1^p$ at small $x$.  However, the accuracy of the data for $x \le 0.05$
is not sufficient.

In this paper, we shall analyze the diffractive pomeron--pomeron cut
contribution to the $g_1(x)$ structure function at small $x$.
This calculation is based on the study of the double--spin
asymmetry in the diffractive $Q \bar Q$ production in the $lp$ reaction
(see Fig.1) which was found not small \ci{all}.  Here, we shall calculate
the planar graphs where the pomeron couples with one quark in the loop.
Then, the quark--pomeron vertex has a simple matrix structure \ci{lana}

\be
 V^\mu_{Pqq} \sim \gamma^\mu. 		\lab{vmu}
\ee
So we can conclude that the effects of the box diagram in Fig.1 are
similar to the contribution of the axial anomaly to $g_1$ \ci{axial,axial1}.

The diffractive part of the structure function $g_1(x,Q^2)$ can be
determined from the spin parallel and spin antiparallel cross--section
difference:
\be
\frac{d^4\sigma(^{\rightarrow} _{\Leftarrow})}{dx dy dx_p dt}
-\frac{d^4\sigma(^{\rightarrow} _{\Rightarrow})}{dx dy dx_p dt}
=\frac{8 \pi \alpha^2}{Q^2} (2-y) g^D_1(x,Q^2,x_p,t).   \lab{g1def}
\ee
Here we neglect the proton mass.
This form differs from the usual $g_1(x,Q^2)$ structure function
definition \ci{ans} only by the additional $x_p$ and $t$ dependence.
The standard set of kinematic variables looks like
\ba
Q^2=-q^2,\;t=(p-p')^2, \nonumber
\\  y=\frac{pq}{p_l p},\;x=\frac{Q^2}{2pq},\;
x_p=\frac{q(p-p')}{qp},\;\beta=\frac{x}{x_p},
\ea
where $p_l,p'_l$ and $p, p'$ are initial and final lepton and proton
momenta, respectively, $q=p_l-p'_l$.

The diffractive structure function $g^D_1$ is determined by the diagram of
Fig.1. The light-quark contribution to $g^D_1$ is
\ba
g^D_1(x,Q^2,x_p,t) =\frac{9 \beta_0^{4} F(t)^{2}}{32\pi^{4} x_p}
I(\beta,Q^2,x_p,t),
\lab{g1i}\\
I(\beta,Q^2,x_p,t)=\int_{k^2_0}^{Q^2/4\beta} \frac{d k_\perp^2
N(\beta,k_\perp^2,x_p,t)}
{\sqrt{1-4k_\perp^2\beta/Q^2}(k_\perp^2+M_Q^2)^2}.
\lab{siglp}
\ea
Here  $M_Q$ is the quark mass,
$\beta_0$ is a quark--pomeron coupling constant, $F(t)$
is a pomeron-proton form factor and
 the function $N$ is determined by the trace over the quark loop.
We integrate in (\ref{siglp}) over $k_\perp^2 \ge k^2_0$ to exclude the
nonperturbative region.
The function $N$ in the $x_p \to 0$ limit looks like
\be
 N(\beta,k_\perp^2,t)=t [2 (\beta-1) k_\perp^{2}+2 (\beta+1) M_Q^{2}
 -\beta t]
\ee
and coincides with the relevant
function from Ref. \ci{axial1} for $x \to 0$.
The result of integration over $dk^2$ in (\ref{siglp}) for $x_p=0$ can be
written in the form
\be
I(\beta,Q^2,x_p,t)= t[2 (\beta-1)  \ln (\frac{Q^{2}}
{(M_Q^{2}+k_0^2) \beta})-\frac{\beta t-4 M_Q^{2}}{M_Q^{2}+k_0^2}].
\ee
So we see that the diffractive part of the structure function will has the
logarithmic $Q^2$ dependence.

Expression (\ref{g1i}) has been obtained for the pomeron with
$\alpha_P(t)=1$. For the supercritical pomeron with $ \alpha_{P}(0) \ge 1$
we must replace the simple power $x_p$ with the power
$x_p^{2\alpha_{P}(t)-1}$.  Here $\alpha_{P} (t)$ is the pomeron trajectory
\be
\alpha_{P} (t)=\alpha_{P} (0)+\alpha_{P}' t
\ee
with $\alpha_{P}'=0.25(GeV)^{-2}$. The behaviour like
$1/x_p^{2\alpha_{P}(t)-1}$ as $x \to 0$ of the diffractive cross section
is connected with the pomeron flux factor \ci{pom}.

The diffractive contribution to $g_1$ can be found
from the integrated $g^D_1(x,Q^2,x_p,t)$ structure function
\be
g^D_1(x,Q^2)=\int_{x}^{x_{max}}dx_p\;\int^{0}_{-\infty} \;dt
\;g^D_1(x,Q^2,x_p,t).
\lab{int}
\ee
 The effects studied are important at small $x_p$, and we
determined the upper limit over $x_p$ in (\ref{int}) as $x_{max}=0.1$.

Let us estimate the leading term of the structure function $g_1$ at small
$x$. For this purpose we shall use the exponential form of the proton
form factor in (\ref{g1i})
\be
F(t)=e^{bt} ;\;\;\;\;\; {\rm with} \;\;\;\;\; b=1.9(GeV)^{-2}.
\ee
We have the following  $x_p$ and $t$--dependence of the  diffractive
structure function $g^D_1$ in (\ref{g1i})
\be
 g^D_1(x,Q^2,x_p,t) \sim \frac{t e^{2bt} }{x_p^{2(\alpha_{P}
 (0)+\alpha_{P}'t)-1}}
 =\frac{t e^{2t(b+\alpha_{P}' \ln{1/x_p})}}{x_p^{2\alpha_{P}
 (0)-1}}.
\ee
Integration  over $x_p$ and $t$ in (\ref{int}) gives us the following
low-x behaviour of the $g_1$ structure function caused by the diffractive
contribution
\be
 g^D_1 \propto \frac{1}{x^{2\alpha_{P}(0)-2}(b+\alpha_{P}'
 \ln{1/x})^2}.   \lab{g1x}
 \ee
This form coincides with (\ref{g1c}) for $\alpha_{P}(0) \sim 1.5$ that is
typical of the BFKL pomeron \ci{bfkl}.
However, this pomeron is now in
contradiction  with the diffractive data \ci{diffr} from HERA which
give:
\be
2\alpha_{P}(0)-1=  \begin{array}{l}
 1.2 \pm 0.1 \;\;\; {\rm for \;\;\; H1}\\
 1.3 \pm 0.1 \;\;\; {\rm for \;\;\; ZEUS}.
 \end{array}
\ee
So the HERA diffractive data can be described by the "soft" pomeron
exchange \ci{pom} with the intercept $\alpha_{P}(0)=1.1-1.15$. We shall
use these values in our future analysis.

The numerical calculation of $g^D_1$ (\ref{g1i}) were performed for
$\beta_0=2(GeV)^{-1}$ and $k^2_0=0.5(GeV)^{-2}$. The
results of calculations can be parametrized for $Q^2=10(GeV)^2$ by the
simple form (\ref{g1x}):
\ba
g^D_1(x,\alpha_{P}(0)=1.1) =\frac{7}{x^{0.2}(1.9+0.25 \ln{1/x})^2}-1.81,
\nonumber \\
g^D_1(x,\alpha_{P}(0)=1.15)=\frac{1.6}{x^{0.3}(1.9+0.25 \ln{1/x})^2}-.52.
\lab{g1app}
\ea
Two sorts of curves are shown in Fig.2 for $\alpha_{P}(0)=1.1$ and
$\alpha_{P}(0)=1.15$.  The first curves are the sum of possible constant
low-$x$ contribution and $g^D_1$:
\be
g^{D+const}_1(x)=0.35+g^D_1(x),
\ee
which are a little lower than the SMC experimental data at $x \le 0.01$.
The second curves are the sum of the nonperturbative pomeron
contribution (\ref{g1l}) and diffractive $g^D_1$
\be
g^{D+L}_1(x)=g_1^L(x)+g^D_1(x),
\ee
These curves are consistent with the low-$x$ SMC experimental data.
It can be seen from Fig.2 that we have found the weak $\alpha_{P}(0)$
dependence of $g^D_1$ for $x \ge 10^{-4}$.

The contribution of the obtained singular behaviour of $g_1$
structure function to the $\Gamma_1$ integral (\ref{gam}) is the
following:
\ba
\Delta \Gamma_1(1.1)=\int_{0}^{0.01}\;dx\;g^{D+L}_1(x,\alpha_{P}(0)=1.1)
=0.011;
\nonumber\\
\Delta
\Gamma_1(1.15)=\int_{0}^{0.01}\;dx\;g^{D+L}_1(x,\alpha_{P}(0)=1.15)=0.012.
\lab{small}
\ea
Thus, the small--$x$ contribution to $\Gamma$ is not small.
The effects from the heavy quarks  in the loop in Fig.1 are
about $10-15\%$ of the light quark contribution (\ref{g1app}). The
complicated quark-pomeron vertex structure, different from the form
(\ref{vmu}) introduced in Ref. \ci{gol-pl}, increases $g_1^D$ and $\Delta
\Gamma_1$ by the same small magnitude.

Note that the contribution of the form  (\ref{g1c}) to  $\Delta \Gamma_1$ is
very large
\be
\Delta \Gamma_1^C=\int_{0}^{0.01}\;dx\;g^{C}_1(x)=0.037;
\ee
which is determined by the extreme singular behaviour of $g_1^C$ at $x \to
0$. However, this small--$x$ approximation is not confirmed by the
 QCD--model calculation presented here.

The $Q^2$ evolution of the diffractive
contribution to $g_1^p(x,Q^2)$ in the HERA $Q^2$ domain is shown in
Fig.3.  The structure function remains positive for all $Q^2$.

Thus, it is shown that the diffractive contribution to $g_1$ leads to the
singular behaviour of the spin--dependent structure function at small $x$,
which is compatible with the SMC data on $g_1^p$.  It is well known that
the pomeron is a colour singlet object which interacts identically with
proton and neutron. This means that the small--$x$ diffractive
contribution to the proton and neutron $g_1$ structure function  will
be the same. This conclusion is not yet confirmed by the SMC data on
$g_1^n$ \ci{smcn}. The neutron structure function $g_1^n$ is negative,
which might be determined by the large negative flavour non--singlet
contribution.  Note that the SMC experimental points on $g_1^n$ at $x \le
0.01$ can indicate possible change in the behaviour of this function to
the positive value.

Then, we can conclude that the singular form of
$g_1$ can affect only the Ellis--Juffe sum rule where the contribution of
the small--$x$ region (\ref{small}) can reach $10\%$ of $\Gamma_1$.
The Bjorken sum rule does not change because it has the flavour
non--singlet contributions only.
The  obtained growth of the proton and
possible change in the behaviour of the neutron  $g_1$ structure
functions can be confirmed by the reduction of the experimental errors in
the SMC data.  The small--$x$ region can be studied in future polarized
experiments at HERA where the range up to $x \sim 10^{-4}$ can be tested.
\vspace{1cm}

The author expresses his deep gratitude to A.V.Efremov, V.G.Krivokhizin,
G.Mallot, W.-D.Nowak, A.Sch\"afer, G.I.Smirnov and O.V.Teryaev for fruitful
discussions.  Special thanks to  I.A.Savin who drew my attention to the
problem of the small -$x$ behaviour of $g_1$ and for useful discussions.

\newpage
{\bf Figure captions}\\
{\bf Fig.1} ~Diffractive contribution to the $g_1$ structure function.\\
{\bf Fig.2} ~$g_1^p(x)$ at $Q^2=10(GeV)^2$.
The sum of the constant low-$x$ contribution and diffractive $g^D_1$:
solid line -- for $\alpha_{P}(0)=1.1$, dot-dashed line
 -- for $\alpha_{P}(0)=1.15$.
The sum of the nonperturbative pomeron contribution (\ref{g1l}) and
$g^D_1$:
dashed line -- for $\alpha_{P}(0)=1.1$, dotted line
 -- for $\alpha_{P}(0)=1.15$.
Experimental data are from Ref. \ci{smc}.\\
{\bf Fig.3} ~$Q^2$-dependence of $g_1^p$ for $\alpha_{P}(0)=1.1$ in the
HERA domain:  solid line -- for $Q^2=10(GeV)^2$, dot-dashed line -- for
 $Q^2=100(GeV)^2$, dashed line -- for $Q^2=1000(GeV)^2$, dotted line --
 for $Q^2=10000(GeV)^2$.

\newpage
  \vspace*{.5cm}
 \hspace{4.5cm}
\epsfxsize=12cm
{\epsfbox{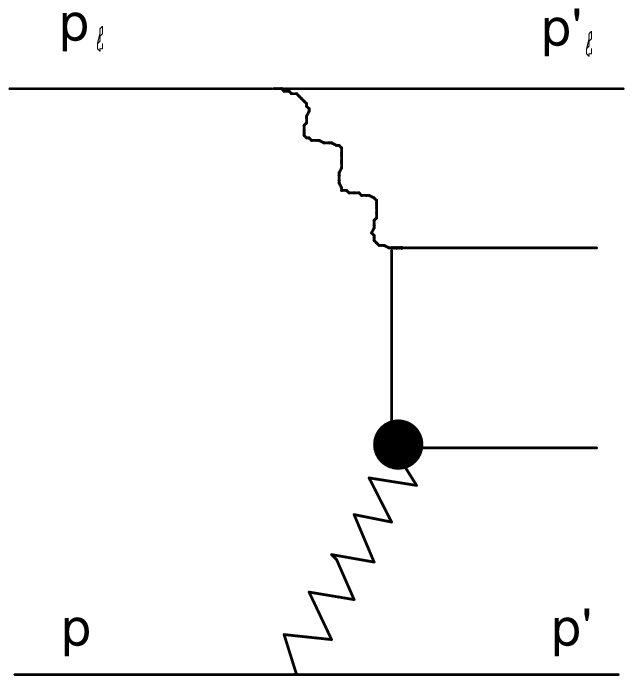}}
  \vspace*{-11.6cm}
\begin{center}
Fig.1
\end{center}

\vspace{1cm}
\samepage
  \vspace*{-.1cm}
\epsfxsize=15cm
\centerline{\epsfbox{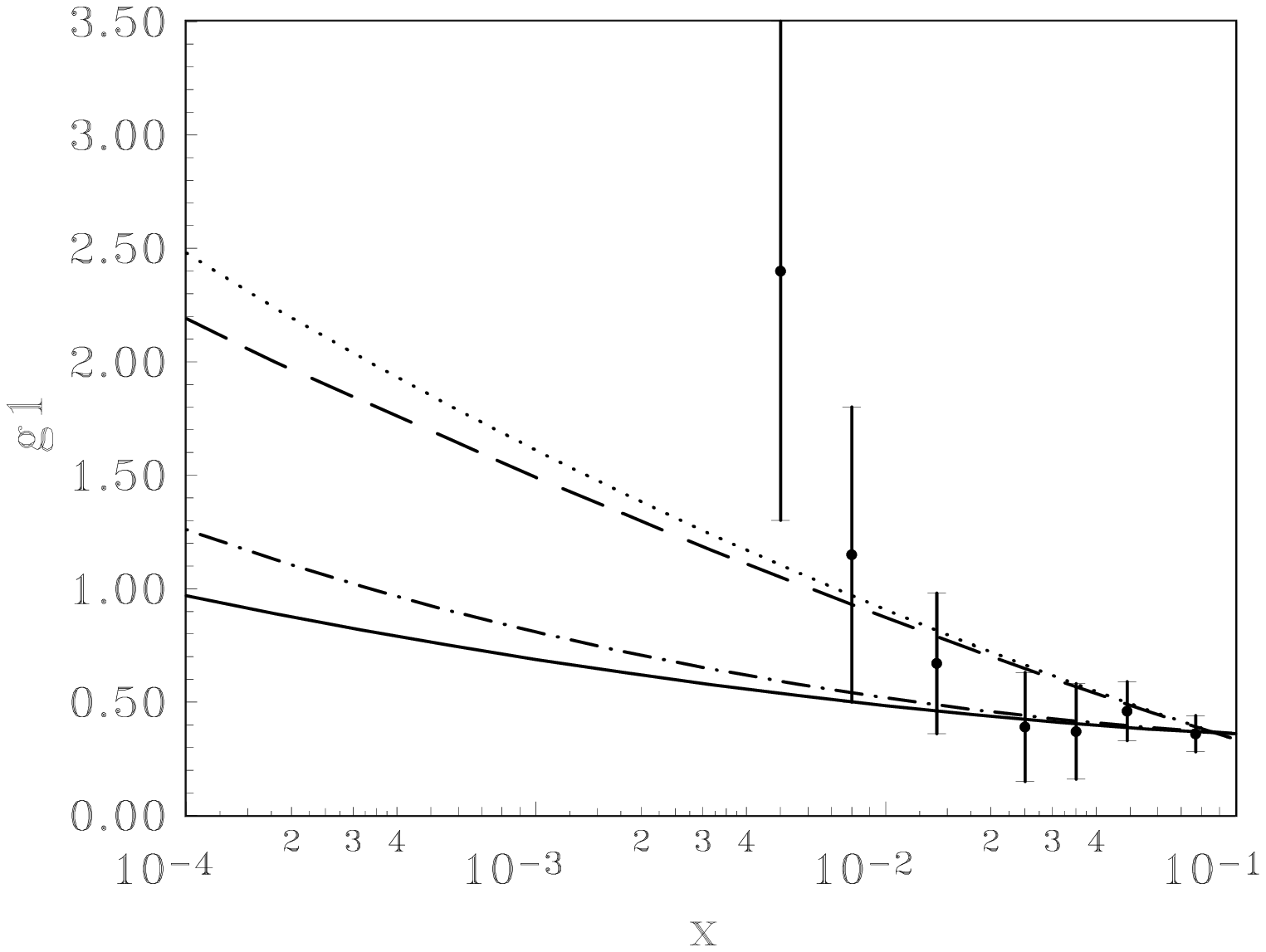}}
  \vspace*{.3cm}
\begin{center}
Fig.2
\end{center}

\newpage
  \vspace*{-.5cm}
\epsfxsize=15cm
\centerline{\epsfbox{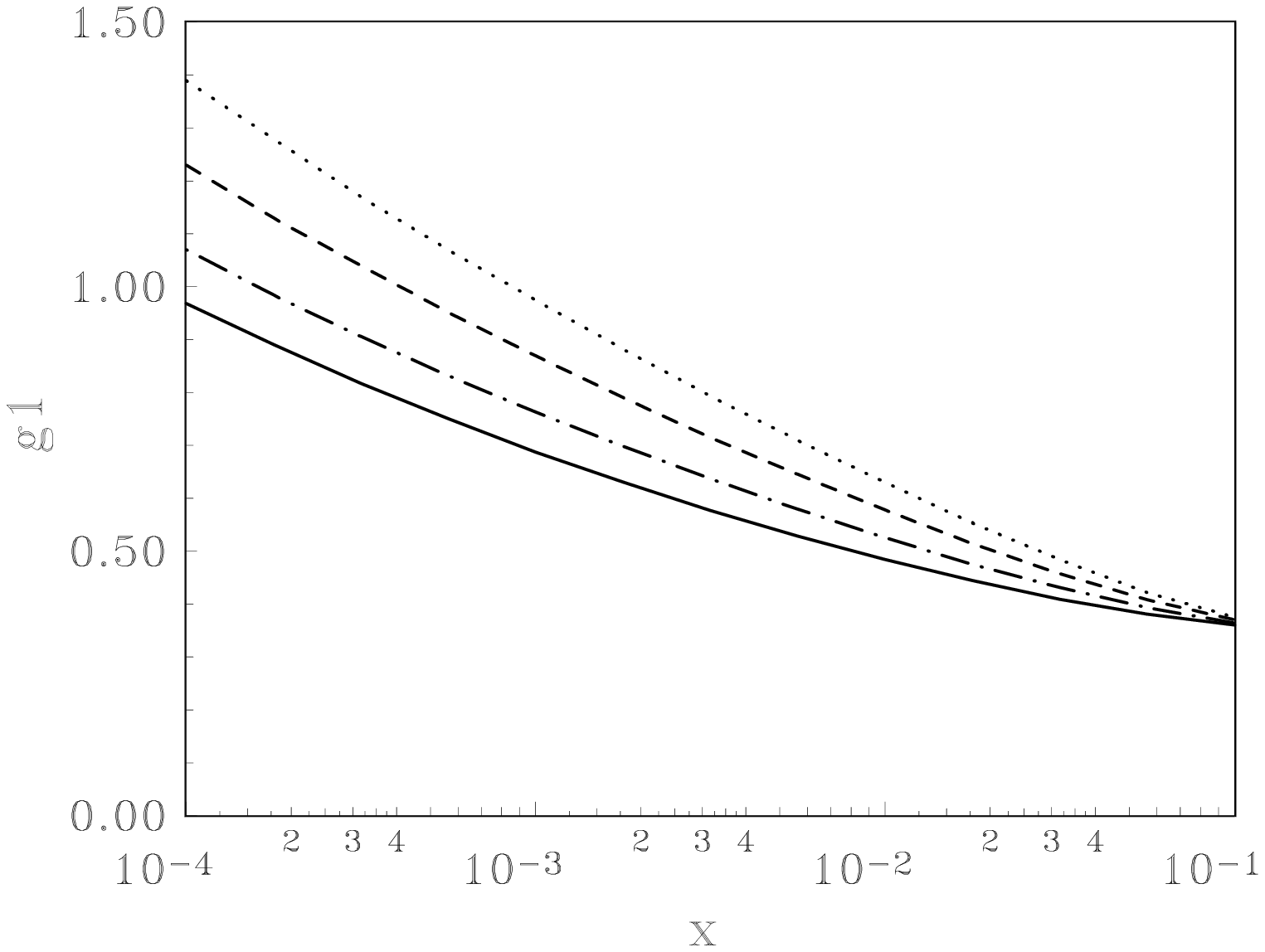}}
  \vspace*{.2cm}
\begin{center}
Fig.3
\end{center}
\end{document}